\RequirePackage{amsmath}
\documentclass[runningheads]{llncs}
\usepackage{graphicx}
\usepackage{tikz}

\usepackage{amsmath}
\usepackage[misc]{ifsym} 
\usepackage{bbding}

\usepackage{wrapfig}
\DeclareMathOperator*{\argmin}{\arg\!\min}
\begin{document}
\title{Adversarial regression training for visualizing the progression of chronic obstructive pulmonary disease with chest x-rays}
\titlerunning{Adversarial regression training for visualizing COPD in chest x-rays}
\author{Ricardo Bigolin Lanfredi\inst{1}\orcidID{0000-0001-8740-5796} \and
Joyce D. Schroeder\inst{2}\orcidID{0000-0002-7451-4886} \and
Clement Vachet\inst{1}\orcidID{0000-0002-8771-1803} \and
Tolga Tasdizen\inst{1}\orcidID{0000-0001-6574-0366}}
\authorrunning{R. Bigolin Lanfredi et al.}
\institute{Scientific Computing and Imaging Institute,\\University of Utah, Salt Lake City UT 84112, USA\\
\email{ricbl@sci.utah.edu} \and
Department of Radiology and Imaging Sciences,\\University of Utah, Salt Lake City UT 84112, USA}
\maketitle
\begin{abstract}
Knowledge of what spatial elements of medical images deep learning methods use as evidence is important for model interpretability, trustiness, and validation. There is a lack of such techniques for models in regression tasks. We propose a method, called visualization for regression with a generative adversarial network (VR-GAN), for formulating adversarial training specifically for datasets containing regression target values characterizing disease severity. We use a conditional generative adversarial network where the generator attempts to learn to shift the output of a regressor through creating disease effect maps that are added to the original images. Meanwhile, the regressor is trained to predict the original regression value for the modified images. A model trained with this technique learns to provide visualization for how the image would appear at different stages of the disease. We analyze our method in a dataset of chest x-rays associated with pulmonary function tests, used for diagnosing chronic obstructive pulmonary disease (COPD). For validation, we compute the difference of two registered x-rays of the same patient at different time points and correlate it to the generated disease effect map. The proposed method outperforms a technique based on classification and provides realistic-looking images, making modifications to images following what radiologists usually observe for this disease. Implementation code is available at \url{https://github.com/ricbl/vrgan}.
\end{abstract}

\section{Introduction}
Methods of visual attribution in deep learning applied to computer vision are useful for understanding what regions of an image models are using~\cite{visreview}. These methods improve a model's interpretability, help in building user trust and validate if the model is using the evidence humans would expect it to use for its task. They have been mostly used to explain models in classification tasks. A common way of formulating the problem of visual attribution is by asking ``What regions of the image are weighted positively in the decision to output this class?''. This question is not suitable for regression tasks, for which we propose to ask ``What would this image look like if it had this other regression value?''. 

To answer the proposed question, we draw from conditional generative adversarial networks (GANs)~\cite{cgan} and image-to-image GANs~\cite{im2im}, both of which have been shown to model complex non-linear relations between conditional labels, input images, and generated images. We hypothesize that using the regression target value for training a visual attribution model with the proposed novel loss will improve the visualization when compared to a similar formulation that only uses classification labels since it does not lose the information of a continuous regression target value by imposing a set of classes. We name our method visualization for regression with a generative adversarial network (VR-GAN).

We study our model in the context of how a value characterizing chronic obstructive pulmonary disease (COPD) is related to changes in x-ray images. COPD is defined by pulmonary function tests (PFTs)~\cite{pft}. A PFT measures forced expiratory volume in one second ($FEV_1$), which is the volume of air a patient can exhale in one second, and forced vital capacity ($FVC$), which is the total volume of air a patient can exhale. A patient with $FEV_1/FVC$ ratio lower than 0.7 is diagnosed to have COPD. 

Radiology provides a few clues that can be used for raising suspicion of COPD directly from an x-ray~\cite{copdxray}. There is a higher chance of COPD when diaphragms are low and flat, corresponding to high lung volumes, and when the lung tissue presents low-attenuation (dark, or lucent), corresponding to emphysema, air trapping or vascular pruning. We show that our model highlights low and flat diaphragm and added lucency. Our method is, to the best of our knowledge, the first data-driven approach to model disease effects of COPD on chest x-rays.
\subsection{Related work}
One way to visualize evidence of a class using deep learning is to perform backpropagation of the outputs of a trained classifier~\cite{visreview}. In~\cite{chexnet}, for example, a model is trained to predict the presence of 14 diseases in chest x-rays, and class activation maps~\cite{cam} are used to show what regions of the x-rays have a larger influence on the classifier's decision. However, as shown in~\cite{wgan}, these methods suffer from low resolution or from highlighting limited regions of the original images. 

In~\cite{wgan}, researchers visualize what brain MRIs of patients with mild cognitive impairment would look like if they developed Alzheimer's disease, generating disease effect maps. To solve problems with other visualization methods, they propose an adversarial setup. A generator is trained to modify an input image which fools a discriminator. The modifications the generator outputs are used as visualization of evidence of one class. This setup inspires our method. However, instead of classification labels, we use regression values and a novel loss function.

There have been other works on generating visual attribution for regression. In~\cite{heart}, Seah et al. start by training a GAN on a large dataset of frontal x-rays, and then train an encoder that maps from an x-ray to its latent space vector. Finally, Seah et al. train a small model for regression that receives the latent vector of the images from a smaller dataset and outputs a value which is used for diagnosing congestive heart failure. To interpret their model, they backpropagate through the small regression model, taking steps in the latent space to reach the threshold of diagnosis, and generate the image associated with the new diagnosis.

The loss function we provide for this task is similar to the cost function provided in~\cite{advregression}. Unlike our formulation,~\cite{advregression} models adversarial attackers and defenders in a game theoretic sense and arrives at an optimal solution for the defenders, using only linear models and applying it to simple features datasets. In~\cite{cganregression}, Bazrafkan et al. propose a method for training GANs conditioned in a continuous regression value. However, the model has a discriminator in parallel with the regressor, it is not used for visual attribution, and the used loss function is different than the one we propose. 
\section{Method}
\subsection{Problem Definition}
We want to generate what an image would look like for different levels of a regression target value, without changing the rest of the content of the image. To formalize this mathematically, we can model an image as $x = f(y,z)$, where $x$ is a dependent variable representing an image, $y$ is an independent variable that determines an aspect of $x$, and $z$ another independent variable representing the rest of the content. In our application, $x$ is an x-ray image, $y$ is the value from a PFT of the same patient taken contemporaneously to the chest x-ray approximating the severity of COPD for that patient, and $z$ represents factors such as patient anatomy unrelated to COPD and position of the body at the moment the x-ray was taken.

We want to construct a model that, given an image $x$ associated with a value $y$ and a content $z$, can generate an image $x'$ conditioned on the same $z$, but on a different value $y'$. By doing this, we can visualize what impacts the change of $y$ to $y'$ has on the image. Similar to~\cite{wgan}, we formulate the modified image as
\begin{equation}
x' = \Delta x + x = G(x, y',y) + x,
\end{equation}
where $G$ is a conditional generator, and $\Delta x$ is a difference map or a disease effect map. By summing $\Delta x$ to $x$, the task of $G$ is made easier, since $G$ only has to model the impact of changing $y$ to $y'$, and the content $z$ should be already in $x$.
\subsection{Loss function}
 \begin{figure}
 \centering
\includegraphics[width=\textwidth,trim={1.5cm 1.1cm 1cm 15.8cm},clip]{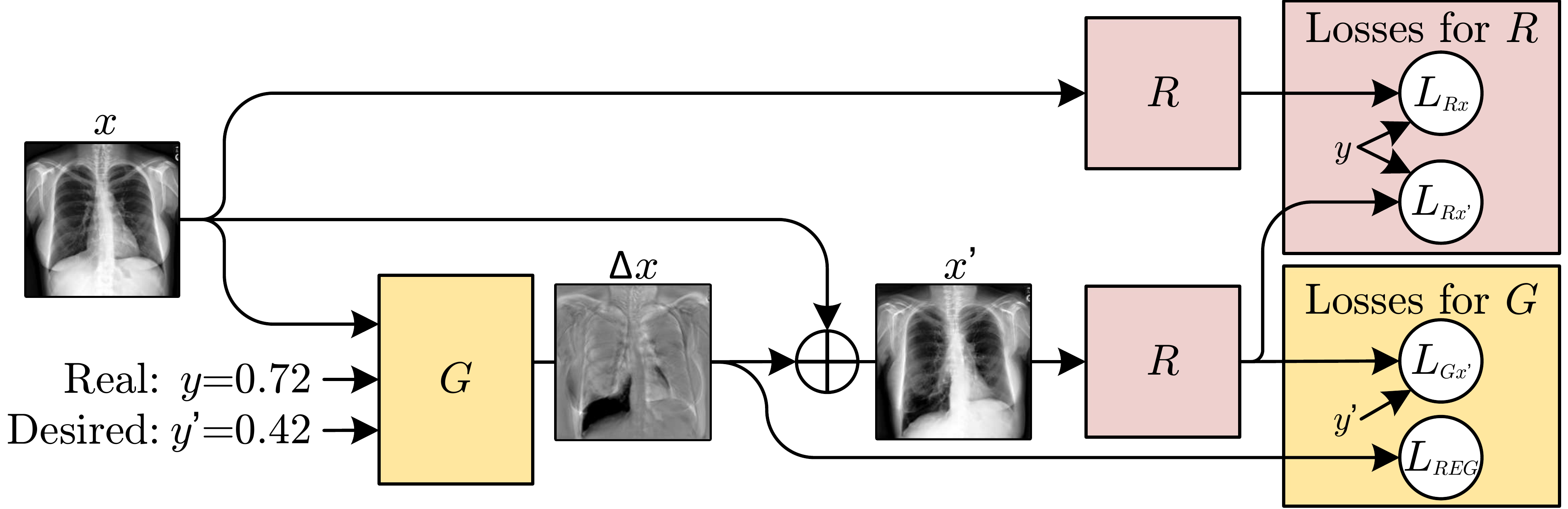}
\caption{Overall model architecture for training with the proposed adversarial loss. The losses $L_{Rx}$, $L_{Rx'}$ and $L_{Gx'}$ are $L1$ regression losses, and $L_{REG}$ is an $L1$ norm penalty. } \label{diagram_RG}
\end{figure}
Fig.~\ref{diagram_RG} shows the loss terms that are defined in this section and how the modules are connected for training a VR-GAN. We start by defining a regressor $R$ that has the task to, given an image $x$, predict its $y$ value. We start building our loss function by defining a term, which is used to optimize only the weights of $R$, to assure $R$ can perform the task of regression over the original dataset: 
\begin{equation}
L_{Rx} = L_{Rx}(x, y)  = \left\lVert R(x) - y \right\rVert_{L1}.
\end{equation}
The regressor is used to assess how close to having $y'$ an image $x'$ is. We also define a term, which is used to optimize only the weights of $G$, to make $G$ learn to create a map $\Delta x$ that, when added to the original image, changes the output of $R$ to a certain value $y'$:
\begin{equation}
\label{gterm}
L_{Gx'} = L_{Gx'}(\Delta x+x, y') = \left\lVert R(G(x, y',y)+x) - y'\right\rVert_{L1}.
\end{equation}
Training a model using only these two terms would lead to an $R$ that does not depend on $G$, and, consequently, to a $G$ that can modify the output of $R$ in simple and unexpressive ways, similar to noisy adversarial examples~\cite{advexamples}. We define an adversarial term, which is used to optimize the weights of $R$, to make $R$ learn to output the same value as the original image for the modified image: 
\begin{equation}
\label{advr}
L_{Rx'} = L_{Rx'}(\Delta x+x, y) = \left\lVert R(G(x, y',y)+x) - y\right\rVert_{L1}.
\end{equation}
As $G$ learns trivial or unrealistic modifications to images, $R$ learns to ignore them due to Eq. (\ref{advr}), forcing the generator to create more meaningful $\Delta x$. This game between $G$ and $R$ should reach an equilibrium where $G$ produces images that are realistic and induces the desired output from $R$. At this point, $R$ should not be able to find modifications to ignore, being unable to output the original $y$. In our formulation, $R$ replaces a discriminator from a traditional GAN.

We define another loss term to assure that $G$ only generates what is needed to modify the label from $y$ to $y'$ and does not modify regions that would alter $z$. An $L1$ penalty is used over the difference map to enforce sparsity:
\begin{equation}
\label{penalty}
L_{REG}=L_{REG}(\Delta x) = \left\lVert \Delta x \right\rVert_{L1}.
\end{equation}
Intuitively, when the modifications generated by $G$ are unrealistic and ignored by $R$, not having any impact to the term defined in Eq. (\ref{gterm}), the norm penalty defined in Eq. (\ref{penalty}) should enforce their removal from the disease effect map. 

The complete optimization problem is defined as 
\begin{equation}
G^* = \argmin_{G} (\lambda_{Gx'} L_{Gx'}+\lambda_{REG} L_{REG}), R^* = \argmin_{R} (\lambda_{Rx'} L_{Rx'}+\lambda_{Rx} L_{Rx}),
\end{equation}
where the $\lambda$'s are hyperparameters. Optimizations are performed alternatingly.
\section{Experiments}
 We used a U-Net~\cite{unet} as $G$. The conditioning inputs $y$ and $y'$, together with their difference, were normalized and concatenated to the U-Net bottleneck layer (Fig.~\ref{diagram_RG}). For $R$, we used a Resnet-18~\cite{resnet}, pretrained on ImageNet and with the output changed to a single linear value. We froze the batch normalization parameters in $R$, as in~\cite{wgan}. Since $R$ depends on the supervision from the original regression task, it will only be able to learn to output values in the range of the original $y$. Therefore, during training we sampled $y'$ from the same distribution as $y$. The hyperparameters were chosen as $\lambda_{Gx'}=0.3$, $\lambda_{REG}=0.03$, $\lambda_{Rx}=1.0$, $\lambda_{Rx'}=0.3$, using validation over the toy dataset presented in Section \ref{ssec:toy}. The same set of hyperparameters were used for the x-ray dataset and were not sensitive to change of dataset. Adam~\cite{adam} was used as the optimizer, with a learning rate of $10^{-4}$. To prevent overfitting, early stopping was used. 
 
 We employed the VA-GAN method presented in~\cite{wgan} as a baseline, since it is a classification version of our method\footnote{Our implementations of VA-GAN and VR-GAN extends code from \url{github.com/orobix/Visual-Feature-Attribution-Using-Wasserstein-GANs-Pytorch}}. We used $\lambda=10^2$ and gradient penalty with a factor of 10, as in~\cite{wgan}. Baseline optimizers and models were the same as the ones described for our model.
 
 We compared the results visually to check if they agreed with radiologists' expectations. For quantitative validation, we used the normalized cross-correlation between the generated $\Delta x$ map and the expected $\Delta x$ map, averaged over the test set. Each result is given with its mean and its standard deviation over 5 tests, with training initialized using distinct random seeds. 
\subsection{Toy dataset}\label{ssec:toy}
\begin{figure}
\begin{minipage}[t]{0.13\textwidth}
  \centering
  \includegraphics[width=1.0\linewidth]{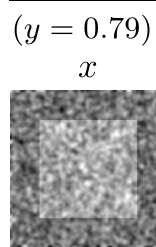}
  (a)
\end{minipage} 
  \centering
\begin{minipage}[t]{0.72\textwidth}
  \centering
  \includegraphics[width=1.0\linewidth]{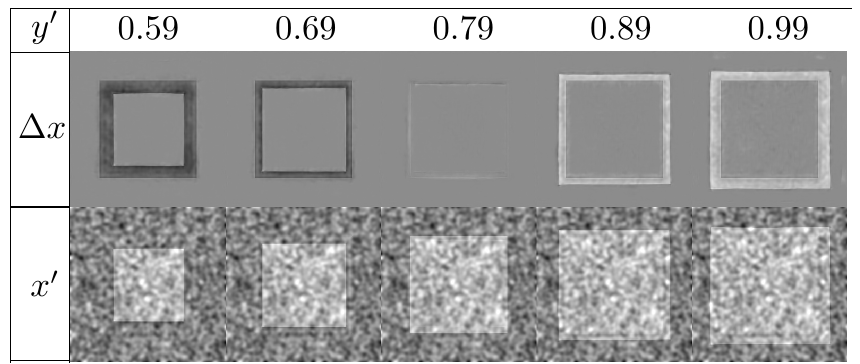}
  (b)
\end{minipage} 
\begin{minipage}[t]{0.13\textwidth}
  \centering
  \includegraphics[width=1.0\linewidth]{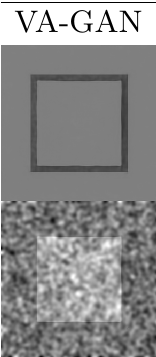}
  (c)
\end{minipage}
\caption{Results on a test example from the toy dataset. \textbf{Top:} difference maps $\Delta x$. \textbf{Bottom:} images of squares, $x$ or $x'$. \textbf{(a)} Original image $x$, with square size $y=0.79$. \textbf{(b)} VR-GAN result for several desired square side lengths $y'$. \textbf{(c)} VA-GAN result.  }
\label{toy_results}
\end{figure}

To test our model, we generated images of squares, superimposed with a Gaussian filtered white noise and with a resolution of $224 \times 224$. An example is presented in Fig.~\ref{toy_results}(a). The side length of the square is proportional to a regression target $y$ that follows a Weibull distribution with a shape parameter of $7$ and a scale parameter of $0.75$. The class threshold for our baseline model~\cite{wgan} was set at $0.7$. It was trained to receive images of big squares ($y\ge0.7$), and output a difference map that made that square small ($y<0.7$). We generated 10,000 images for training. Since we generated the images, we could evaluate with perfect ground truth for $\Delta x$. We evaluated using input examples where $y\ge0.7$ and sampling $y'<0.7$ from the Weibull distribution. This resulted in 5,325 images for validation and 5,424 images for testing. 

Examples of difference maps and modified versions of the original image for a few levels of the desired square side length are presented in Fig.~\ref{toy_results}(b). While the baseline presents a fixed modification for an image, shown in Fig.~\ref{toy_results}(c), and can only generate smaller squares, our method can generate different levels of change for the map, and also generate both bigger and smaller squares. The baseline~\cite{wgan}, using VA-GAN, obtained a score of $0.780\pm0.007$ for the normalized cross-correlation, while our method, VR-GAN obtained a score of $0.853\pm0.014$. 
\subsection{Visualizing the progression of COPD}
\begin{figure}
\begin{minipage}[t]{0.115\textwidth}
  \centering
  \includegraphics[width=1.0\linewidth]{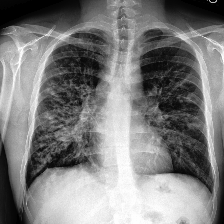}
  (a)
\end{minipage}
\begin{minipage}[t]{0.115\textwidth}
  \centering
  \includegraphics[width=1.0\linewidth]{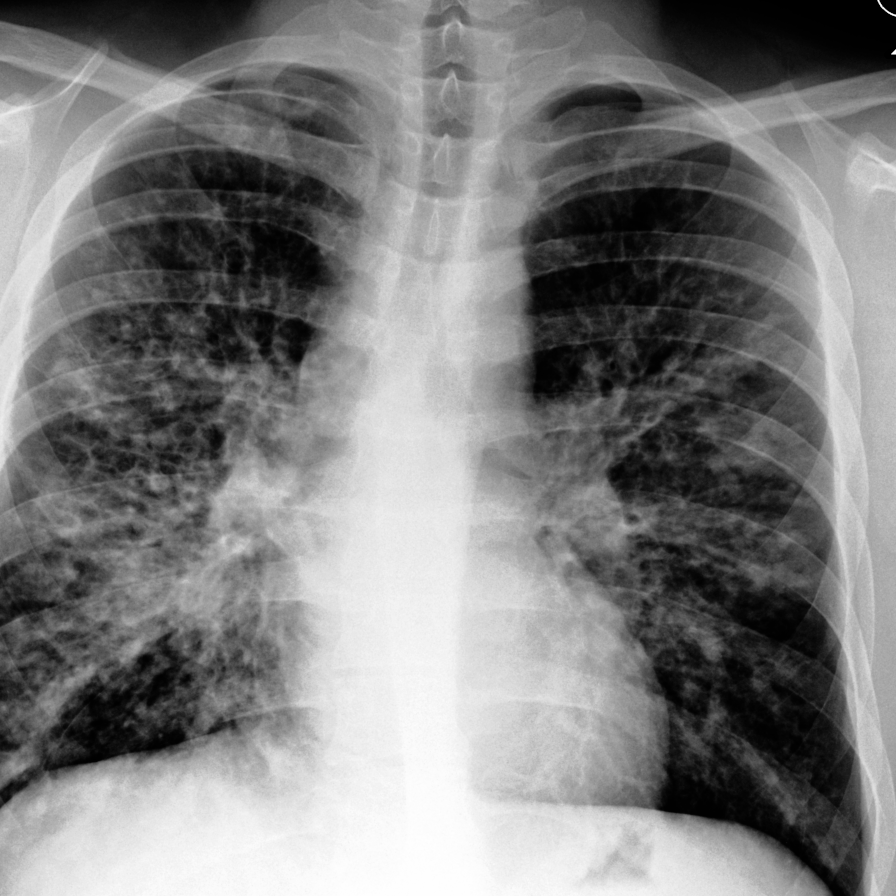}
  (b)
\end{minipage}
\begin{minipage}[t]{0.115\textwidth}
  \centering
  \includegraphics[width=1.0\linewidth]{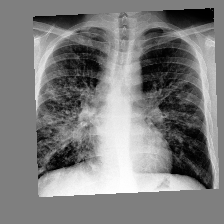}
  (c)
\end{minipage}
\begin{minipage}[t]{0.115\textwidth}
  \centering
  \includegraphics[width=1.0\linewidth]{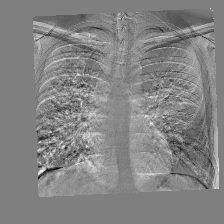}
  (d)
\end{minipage}
\begin{minipage}[t]{0.115\textwidth}
  \centering
  \includegraphics[width=1.0\linewidth]{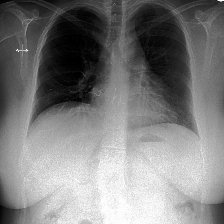}
  (e)
\end{minipage}
\begin{minipage}[t]{0.115\textwidth}
  \centering
  \includegraphics[width=1.0\linewidth]{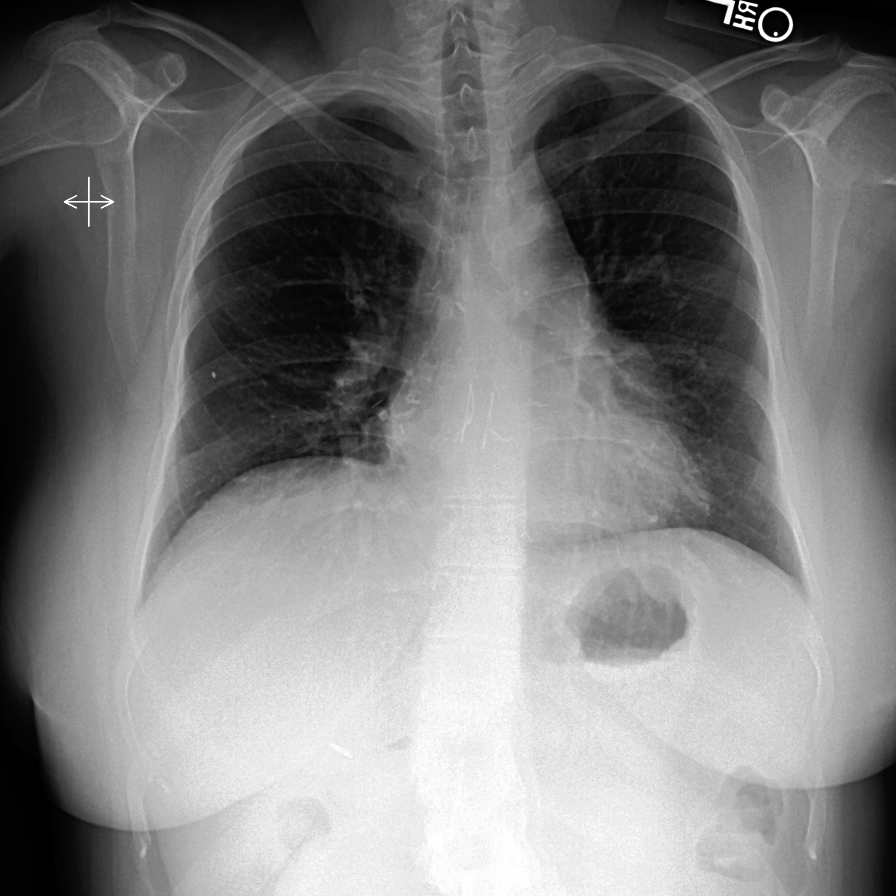}
  (f)
\end{minipage}
\begin{minipage}[t]{0.115\textwidth}
  \centering
  \includegraphics[width=1.0\linewidth]{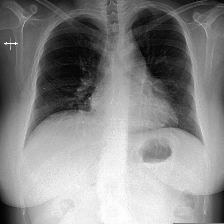}
  (g)
\end{minipage}
\begin{minipage}[t]{0.115\textwidth}
  \centering
  \includegraphics[width=1.0\linewidth]{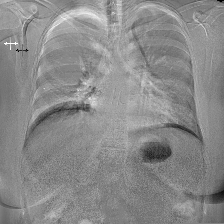}
  (h)
\end{minipage}
\caption{Examples of results of the alignment. \textbf{(a) and (e):} Reference images without COPD. \textbf{(b) and (f):} Images to align, with COPD. \textbf{(c) and (g):} Aligned images. \textbf{(d) and (h):} Difference between reference and aligned images, used as $\Delta x$ ground truth.} \label{registration}
\begin{minipage}[t]{0.13\textwidth}
  \centering
  \includegraphics[width=1.0\linewidth]{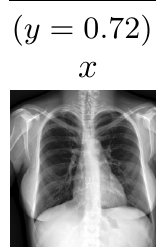}
  (a)
\end{minipage}
\centering
\begin{minipage}[t]{0.72\textwidth}
  \centering
\includegraphics[width=1.0\linewidth]{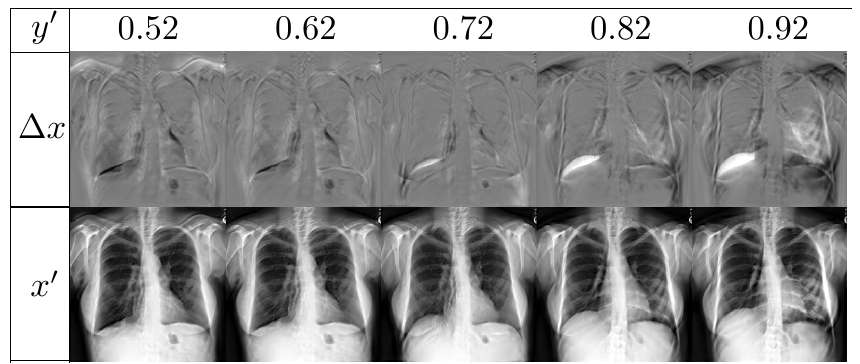}
  (b)
\end{minipage} 
\begin{minipage}[t]{0.13\textwidth}
  \centering
\includegraphics[width=1.0\linewidth]{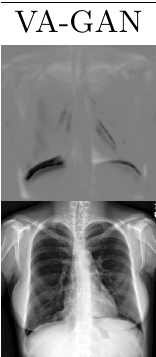}
  (c)
\end{minipage}
\caption{Results on a test example of the COPD dataset. \textbf{Top:} disease effect maps $\Delta x$. \textbf{Bottom:} chest x-rays, $x$ or $x'$. \textbf{(a)} Original image $x$, with $FEV_1/FVC$ ($y$) 0.72. \textbf{(b)} VR-GAN results for several desired $FEV_1/FVC$ ($y'$). The lower this value, the more severe the disease. \textbf{(c)} VA-GAN results.  }
\label{copd_results} 
 \centering
\begin{minipage}[t]{0.885\textwidth}
  \centering
 \includegraphics[width=1.0\linewidth]{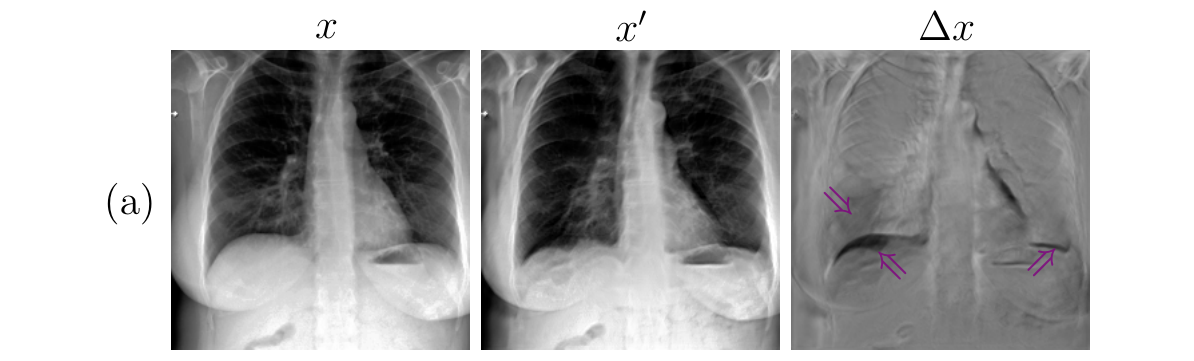}
\end{minipage}
\centering
\begin{minipage}[t]{0.885\textwidth}
  \centering
 \includegraphics[width=1.0\linewidth]{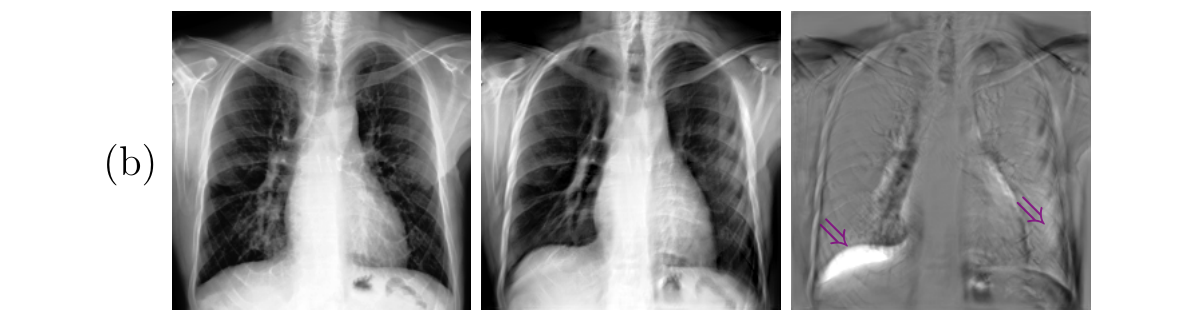}
\end{minipage}
\caption{Examples of disease effects that correspond with visual feature changes expected radiologically. In $\Delta x$, gray represents no change, black a decrease and white an increase in image intensity. From left to right: original image $x$, modified image $x'$ and disease effect map $\Delta x$. \textbf{(a)} $y=0.8$ to $y'=0.4$ (increasing severity). Purple arrows highlight flat and low diaphragm (bottom two arrows) and added lucency (top left arrow). \textbf{(b)} $y=0.37$ to $y'=0.8$ (decreasing severity). Purple arrows highlight high and curved diaphragm (left arrow) and reduced lucency (right arrow).} \label{evidence}
\end{figure}
We gathered a dataset of patients that had a chest x-ray exam and a PFT within 30 days of each other at the University of Utah Hospital from 2012 to 2017. This study was performed under an approved Institutional Review Board process\footnote{IRB\_00104019, PI: Schroeder MD} by our institution. Data was transferred from the hospital PACS system to a HIPAA-compliant protected environment. Orthanc\footnote{\url{orthanc-server.com}} was used for data de-identification by removing protected health information. Lung transplants patients were excluded, and only posterioranterior (PA) x-rays were used. PFTs were only associated with their closest x-ray exam and vice-versa. For validation and testing, all subjects with at least one case without COPD (used as original image $x$) and one case with COPD (used as desired modified image $x'$) were selected, using COPD presence as defined by PFTs. This setup was chosen because the trained baseline model can only handle transitions from no disease to disease. For each of these subjects, we used all combinations of pair of cases with distinct diagnoses. The average time between paired exams was 17 months. We used 3,414 images for training, 208 pair of images for validation and 587 pair of images for testing. Images were cropped to a centered square, resized to $256 \times 256$ and randomly cropped to $224 \times 224$. We equalized their histogram and normalized their individual intensity range to [-1,1]. We used $FEV_1/FVC$ as the regression target value $y$. To generate ground truth disease effect maps, we aligned two x-rays of the same patient, using an affine registration employing the pystackreg library\footnote{bitbucket.org/glichtner/pystackreg}, and subtracted them. Examples are shown in Fig.~\ref{registration}.

Disease effect maps and modified images for both methods are presented in Fig.~\ref{copd_results}. Note that our method can modify images to increase and decrease severity by any desired amount in contrast to~\cite{wgan}, which can only be trained to modify the classification of images in a single direction. In Fig.~\ref{evidence}, we show images generated using VR-GAN which highlight the height and flatness of diaphragm and show changes in the level of lung lucency, features that radiologists use as evidence for COPD on chest x-rays. Small changes in the cardiac contour are consistent with the accommodation of a shift in lung volume. Using normalized cross-correlation, VA-GAN obtained a score of 0.012$\pm$0.015, while VR-GAN obtained a score of 0.127$\pm$0.017. The low correlation scores may result from imperfect alignments with affine transformations and potential changes between x-ray pairs unrelated to COPD. However, our method still obtained a significantly better score.

\section{Conclusion}
We introduced a visual attribution method for datasets with regression target values and validated it for a toy task and for chest x-rays associated with PFTs, assessing the impact of COPD in the images. We demonstrated significant improvement in the disease effect maps generated by a model trained with adversarial regression when compared to a baseline trained using classification labels. Furthermore, the generated disease effect maps highlighted regions that agree with radiologists' expectations and produced realistic images.

\end{document}